\documentclass[aps,prb,twocolumn,showpacs,floatfix]{revtex4}
\usepackage{amsmath,amssymb,amsfonts}
\usepackage{graphicx}
\usepackage{hyperref}
\begin{document}

\title{Investigation on Anharmonicity, Vibrational Anisotropy and Thermal Expansion of an Amorphous Ni$_{46}$Ti$_{54}$
Alloy Produced by Mechanical Alloying using Extended X-ray Absorption Fine Structure}


\author{K. D. Machado}
\email{kleber@fisica.ufpr.br}
\affiliation{Departamento de F\'{\i}sica, Centro Polit\'ecnico, Universidade Federal do Paran\'a, 81531-990,
Curitiba, PR, Brazil}

\date{\today}

\begin{abstract}
A method to investigate anharmonicity, vibrational anisotropy and thermal expansion using correlated
mean-square relative displacements (MSRD) parallel and perpendicular to the interatomic bonds obtained only from
Extended X-ray Absorption Fine Structure (EXAFS) analysis based on cumulant expansion is suggested and applied to
an amorphous Ni$_{46}$Ti$_{54}$ alloy produced by mechanical
alloying. From EXAFS measurements taken on Ni and Ti K edges at several temperatures, the thermal behavior of
$\text{MSRD}_\parallel$, $\text{MSRD}_\perp$ and of the cumulants $C_1^*$, $C_2^*$ and $C_3^*$ of the
real distribution functions $\varrho_{ij}(r,T)$, and also the Einstein temperatures and frequencies associated with
parallel and perpendicular motion were obtained, furnishing information about the anharmonicity of the interatomic
potential, vibrational anisotropy and the contribution of the perpendicular motion to the total disorder and thermal
expansion.

\end{abstract}

\pacs{61.05.cj, 63.20.Ry, 65.60.+a, 63.50.Lm}



\maketitle

\section{Introduction}

Many physical and chemical properties of a material depend on its atomic structure. Thus, in order to understand these
properties, structural studies on crystalline and amorphous materials should be carried on. These studies depend on the
use of techniques able to extract structural information such as average coordination numbers, average interatomic
distances, structural and thermal disorders and so on. In this case, the first approach usually
is to use x-ray diffraction (XRD)
\cite{Cullity,Warren,Guinier}, which is the most common structural technique. XRD can be used mainly for crystals to find
lattice parameters, crystallographic thermal expansion, uncorrelated mean-square displacements (MSD) and relative
quantities of phases present in a sample \cite{klecoge,kleCuSe}.
However, XRD is a long-range probe, and information about a specific atomic
species is not easy to obtain. When the material under study is amorphous, the difficulties increase, and determination
of structural parameters for binary or multicomponent alloys usually require the
use of experimental methods such as anomalous x-ray diffraction (AXRD), neutron diffraction (ND) with isotopic substitution
(IS), or theoretical
methods as molecular dynamics (MD), Monte Carlo (MC) or Reverse Monte Carlo (RMC) simulations, and sometimes a combination
of two or more techniques. Unfortunately, more detailed studies, including investigations on thermal
expansion, anharmonicity of interatomic potentials and vibrational anisotropy, are very difficult to be done on
amorphous alloys using the cited techniques, which leads us to the extended x-ray absorption fine structure spectroscopy
(EXAFS) technique. EXAFS is a powerful tool for obtaining the local atomic order around a specific  atomic
species \cite{Koningsberger,Teo} due to its selectivity. EXAFS oscillations $\chi(k)$ obtained on an edge of an element $A$
furnish information about interactions involving only the element $A$, and the procedure used to extract such information
is almost the same for crystalline or amorphous samples \cite{KleGaSe,kleCoTi}.
In this aspect, due to the high values of $2k$ probed by EXAFS,
valuable information about the  medium and mainly the short-range order can be obtained, a property very relevant for
amorphous alloys since they basically exhibit structures only with this kind of order.
In addition, studies on thermal properties are relatively easy to be carried on if high-quality EXAFS measurements at
different temperatures were taken, which opens the possibility of more sofisticated
investigations.  For moderately disordered systems \cite{Dalba,Tranquada}, the use of
the cumulant expansion analysis \cite{Bunker,Koningsberger2} can take into account thermal effects and also
anharmonicity. Many investigations based on this method have been done since the proposal of EXAFS as
a vibrational probe in the 1970s \cite{Beni,Sevillano}, and
information about static (or structural) disorder, thermal disorder, thermal expansion, anharmonicity effects and
vibrational anisotropy were obtained \cite{Dalba3,Ikemoto,Araujo,kleberse90s10,Schnohr,Purans}.  The main point is
that EXAFS is sensitive to the parallel and perpendicular correlated mean-square relative displacement
\cite{Lee,Dalba4} (MSRD$_\parallel$ and MSRD$_\perp$, respectively) and to the asymmetry of the
one-dimensional effective pair potential \cite{Frenkel,Yokoyama}, and this feature can be exploited to furnish structural
and thermal information related to the alloys under study. Some crystalline alloys were studied considering this approach
but to our knowledge $\text{MSRD}_\perp$ and quantities related to it were not found for any amorphous samples. The main
problem in this case is the determination of the quantity equivalent to the crystallographic distance $R_c$, which in
principle is needed to obtain $\text{MSRD}_\perp$. For crystalline materials, $R_c$ is related to the lattice parameters
and can be obtained, for instance, from a Rietveld refinement procedure \cite{Rietveld2,Rietveld}. For amorphous
samples, on the other hand, it is not easy to determine the amorphous XRD distance $R_a$ and, when it can be done,
usually error bars are large and do not allow reliable values for $\text{MSRD}_\perp$. We developed a method to
extract several properties such as vibrational anisotropy, anharmonicity, asymmetry of distribution functions and thermal
expansion from EXAFS measurements only, and a detailed explanation of the procedure is given below. We illustrate it
by investigating the structural properties of an amorphous Ni$_{46}$Ti$_{54}$ ({\em a}-Ni$_{46}$Ti$_{54}$) alloy produced by
mechanical alloying \cite{MASuryanarayana}. NiTi alloys are very interesting since they can
exhibit shape memory and superelastic effects, excellent ductibility and good fatigue life, good corrosion resistance
and biocompatibility \cite{Shabalovskaya,Duerig,Otsuka,Ju,McIntosh}, being
candidates to use as artificial bones or teeth roots \cite{Lipscomb}. Other applications
include the use of shape memory and superelastic NiTi bars and wires as structural elements in buildings \cite{DesRoches}.
Amorphous NiTi alloys can be produced in a wide compositional range which extends from 20\% to about 70\% Ni
\cite{Buschow}, making this system very suitable for amorphization studies \cite{KleNiTi,nitikleber,kleni46ti54}.
Considering EXAFS measurements on edges of
Ni and Ti at several temperatures, we obtained, besides average coordination numbers
and interatomic distances, $\text{MSRD}_\parallel$ and $\text{MSRD}_\perp$, structural and thermal disorder,
anharmonicity of the effective interatomic pair potential, vibrational anisotropy and thermal expansion considering
correlated Einstein models for the temperature dependence of cumulants $C_2^*$ and $C_3^*$.

The structure of this article is as follows. Sec. \ref{secteoria} presents the theoretical fundamentals needed to the
EXAFS analysis and cumulant expansion. Sec. \ref{secexperimental} shows the experimental procedures used to produce the
alloy and to obtain the EXAFS measurements. Results and detailed discussions are given in sec. \ref{secresultados}, and
sec. \ref{secconclusoes} summarize the conclusions obtained.

\section{Theoretical Background}
\label{secteoria}

To obtain structural information from EXAFS we considered the well known cumulant expansion method
\cite{Bunker,Koningsberger,Fornasini,Dalba4} which is valid for small to moderate disorder. The EXAFS signal on a K
absorbing edge for a coordination shell $\ell$ of an absorbing atom of type $i$ and a backscatter of type $j$ can be
written as \cite{Fornasini,Dalba4,Koningsberger,Bouldin}

\begin{widetext}
\begin{equation}
\chi_\ell^{ij}(k,T) = \frac{S_0^2 N_{ij}}{k} \text{Im} \biggl[f_j(k) e^{2i \delta(k)}\int_0^\infty{\varrho_{ij} (r,T)
\frac{e^{-2r/\lambda}}{r^2} e^{2ikr} \, dr}\biggr] \,,
\label{chi}
\end{equation}

\end{widetext}

\noindent where $S_0^2$ is the amplitude factor associated with intrinsic process that contribute to the photoabsorption
but not to EXAFS, $N_{ij}$ is the average coordination number of atoms of type $j$ around atoms of type $i$,
$f_j(k)$ is the complex backscattering factor, $\delta(k)$ is the phaseshift associated with the absorbing atom,
$\varrho_{ij}(r,T)$ is the partial radial distribution function (RDF) \cite{Bouldin}, $\lambda$ is the
photoelectron mean free path and $T$ is the temperature. The RDF is the {\em real} distribution of distances, and
the {\em effective} distribution of distances is given by

\begin{equation}
\Upsilon_{ij}(r,T, \lambda) = \varrho_{ij} (r,T) \frac{e^{-2r/\lambda}}{r^2}\,.
\label{effective}
\end{equation}

It is important to note that the instantaneous interatomic distances $r_{ij}$ are distributed
according to the real distribution $\varrho_{ij}(r,T)$, but in an EXAFS measurement the photoelectrons,
which have a mean free path $\lambda$, probe the effective distribution $\Upsilon_{ij}(r,T,\lambda)$ due to the
spherical photoelectron wave, to the weakening of this wave with $r$ and to the finite mean free path
\cite{Bunker}. The function

\begin{equation}
\Xi_{ij}(k,T,\lambda) = \int_0^\infty{\varrho_{ij} (r,T) \frac{e^{-2r/\lambda}}{r^2} e^{2ikr} \, dr}
\label{caracteristica}
\end{equation}

\noindent is called the characteristic function \cite {Reichl} associated with the distribution
$\Upsilon_{ij}(r,T,\lambda)$. This function can be though of as being the average value of $e^{2ikr}$, which can be
expanded in a power series of the moments $\langle r^n \rangle$ as

\begin{widetext}
\begin{equation}
\Xi_{ij}(k,T,\lambda) = \langle e^{2ikr} \rangle = \biggl\langle \Bigl(1 + 2ikr + \frac{(2ik)^2r^2}{2!}
+ \frac{(2ik)^3r^3}{3!} + \cdots \Bigr)\biggr\rangle
= \langle 1  \rangle + 2ik \langle r \rangle + \frac{(2ik)^2}{2!} \langle r^2 \rangle
+ \frac{(2ik)^3}{3!} \langle r^3 \rangle + \cdots \,.
\label{xiserie}
\end{equation}

\end{widetext}

The characteristic function can also be written in terms of cumulants $C_n$ through

\begin{multline}
\Xi_{ij}(k,T,\lambda) = \exp{\Bigl[ \sum_{n=1}^{\infty}{\frac{(2ik)^n}{n!} C_n}\Bigr]} \\=
\exp\Bigl[2ik C_1 + \frac{(2ik)^2 C_2}{2!}  + \frac{(2ik)^3 C_3}{3!}  + \cdots\Bigr] \,.
\label{cumulantes}
\end{multline}

\noindent Expanding the right hand side of eq.~\ref{cumulantes} and collecting terms of same order in $2ik$
results in

\begin{multline}
\Xi_{ij}(k,T,\lambda) = 1+  2ik C_1 +
\frac{C_2 +C_1^2}{2!} (2ik)^2
\\
+ \frac{C_3 +3C_1 C_2+ C_1^3  }{3!} (2ik)^3
+ \cdots\,,
\label{cumulantes2}
\end{multline}

\noindent and, comparing eqs.~\ref{xiserie} and \ref{cumulantes2}, the cumulants are given by

\begin{subequations}
\label{cumulantes3}
\begin{align}
C_1 &= \langle r \rangle
\label{cumulantes3a}\\
C_2 &= \langle r^2 \rangle - \langle r \rangle^2 = \langle (r - \langle r \rangle)^2\rangle
\label{cumulantes3b}\\
C_3 &= \langle r^3 \rangle - 3 \langle r \rangle\langle r^2 \rangle + 2\langle r \rangle^3 =
\langle (r  - \langle r \rangle)^3 \rangle
\label{cumulantes3c}\\
\vdots & \notag\\
C_n &=  \langle (r - \langle r \rangle)^n \rangle \,, n \ge 2\,.
\label{cumulantes3d}
\end{align}
\end{subequations}

\noindent These are the cumulants of the effective distribution $\Upsilon(r,T,\lambda)$. When the real distribution

\begin{equation}
g_{ij}(r,T) = \frac{\varrho_{ij}(r,T)}{r^2}
\label{g}
\end{equation}

\noindent is considered, the real cumulants $C_n^*$ are obtained, and the corresponding characteristic function
is

\begin{equation}
\Lambda(k,T) = \int_0^\infty{ \frac{\varrho_{ij}(r,T)}{r^2} \, dr}\,.
\end{equation}

\noindent If this function is expanded, equations similar to eqs.~\ref{xiserie},  \ref{cumulantes} and
\ref{cumulantes3} are found but with cumulants $C_n^*$ instead of $C_n$. It is important to note that to 
analyze anharmonicity, asymmetries, vibrational anisotropy and thermal expansion
we need the cumulants $C_n^*$. In particular, $C_1^*$ is the average interatomic distance, $C_2^*$ is related to the
disorder and $C_3^*$ measures the asymmetry of the distribution function $g_{ij}(r,T)$. Some important quantities
can be obtained directly from these three cumulants.
To see that, let $\vec{u}_j$ and $\vec{u}_0$ be the instantaneous
displacements of the backscatterer and absorber atom, respectively. Defining $\Delta \vec{u} = \vec{u}_j - \vec{u}_0$
as the instantaneous relative thermal displacement between the backscatterer and absorber atoms, the total MSRD is
given by $\text{MSRD} = \langle (\Delta \vec{u})^2 \rangle$, which can be decomposed in a MSRD parallel to the
interatomic bond (MSRD$_\parallel = \langle(\Delta u_\parallel)^2\rangle$) and in a perpendicular one (MSRD$_\perp =
\langle (\Delta u_\perp)^2\rangle$). If $\vec{R}_0$ is the relative position of the backscatter in the absence of
thermal vibrations, the instantaneous relative position is

\begin{equation*}
\vec{r} = \vec{R}_0 + \Delta \vec{u}\,,
\end{equation*}

\noindent and the instantaneous relative distance is \cite{Fornasini,Dalba3}

\begin{equation}
r \simeq R_0 + \Delta u_\parallel + \frac{(\Delta u_\perp)^2}{2R_0}\,,
\label{distanciar}
\end{equation}

\noindent where

\begin{equation*}
\Delta u_\parallel = \hat{R}_0 \cdot \Delta \vec{u} = \hat{R}_0 \cdot (\vec{u}_j - \vec{u}_0)\,.
\end{equation*}

Then, the first cumulant $C_1^*$ is \cite{Dalba,Dalba2,Fornasini,Fornasini2}

\begin{equation}
C_1^* = \langle r \rangle \simeq R_0 + \langle\Delta u_\parallel\rangle + \frac{\langle (\Delta u_\perp)^2\rangle}{2R_0}
\,,
\label{cumulantec1}
\end{equation}

\noindent where

\begin{align}
r_\parallel &= \langle\Delta u_\parallel\rangle \,,& r_\perp &= \frac{\langle (\Delta u_\perp)^2\rangle}{2R_0}\,.
\label{rparaperp}
\end{align}

In a crystalline sample, the crystallographic distance $R_c$ is related to the lattice parameters and can be
obtained, for instance, from a Rietveld refinement procedure \cite{Rietveld,Rietveld2}, and it is given by

\begin{equation}
R_c = R_0 + \langle\Delta u_\parallel\rangle
\label{eqdefrc}
\end{equation}

\noindent In this case, eq. \ref{cumulantec1} becomes

\begin{equation}
C_1^* \simeq R_c +  \frac{\langle (\Delta u_\perp)^2\rangle}{2R_0}
\label{eqrc}
\end{equation}

\noindent Then, in principle, information about $\text{MSRD}_\perp $ can be obtained if
$C_1^*$ and $R_c$ were known from EXAFS and XRD, respectively. The second cumulant is, to first order,

\begin{widetext}
\begin{equation}
C_2^* =  \bigl\langle (r - \langle r\rangle)^2 \bigr\rangle  \simeq \langle (\Delta u_\parallel)^2\rangle =
\text{MSRD}_\parallel
 = \langle [\hat{R}_0 \cdot (\vec{u}_j - \vec{u}_0)]^2\rangle =
\langle (\hat{R}_0 \cdot \vec{u}_j)^2 \rangle + \langle (\hat{R}_0 \cdot \vec{u}_0)^2 \rangle
-2 \langle (\hat{R}_0 \cdot \vec{u}_j) (\hat{R}_0 \cdot \vec{u}_0)\rangle
\end{equation}

\end{widetext}

\noindent
The terms $\langle (\hat{R}_0 \cdot \vec{u}_j)^2 \rangle$ and $\langle (\hat{R}_0 \cdot \vec{u}_0)^2 \rangle$
are the uncorrelated mean square displacements of the backscatterer and absorber atoms, respectively, and can be
obtained from XRD measurements (for crystalline samples). The factor $\langle (\hat{R}_0 \cdot \vec{u}_j)
(\hat{R}_0 \cdot \vec{u}_0)\rangle$ is the displacement correlation function (DCF) \cite{Beni,Dalba4}, and XRD is not
sensitive to it.
An isotropic Debye crystal \cite{Dalba3,Fornasini3} has $\text{MSRD}_\perp = 2 \text{MSRD}_\parallel$, so
if both quantities can be found, it is possible to extract information about anisotropic vibrations, as was done
recently for Cu \cite{Dalba3} and InP \cite{Schnohr}. The third cumulant, given by $C_3^*= \langle (r - \langle r\rangle)^3
\rangle$, measures the asymmetry of the {\em real} unidimensional distribution $\varrho(r,T)$ and can be
associated with the anharmonicity of an {\em effective} interatomic potential

\begin{equation}
V(r) \simeq k_e(r-r_0)^2 - k_3 (r-r_0)^3 \,,
\label{potencialefetivo}
\end{equation}

\noindent where $r_0$ is the minimum of $V(r)$, $k_e$ is the effective harmonic spring constant, and
$k_3$ is the cubic anharmonicity constant.

It is usual to consider temperature dependences for $C_2^*$ and $C_3^*$ based on Einstein
or Debye models \cite{Sevillano,Beni,Vaccari,Dalba4} and, considering the correlated Einstein model,
$C_2^*$ is given through

\begin{equation}
C_2^* = C_{2,\parallel,T}^* +C^*_{2,{\text{st}},\parallel}=
\frac{\hbar^2}{2\mu k_B \Theta_{\parallel}} \coth{\bigl(\frac{\Theta_{\parallel}}{2T} \bigr)} +
C^*_{2,{\text{st}},\parallel} \,,
\label{eqc2}
\end{equation}

\noindent where $h = 2 \pi \hbar$ is the Planck\textquoteright s constant, $\Theta_\parallel$ is the Einstein
temperature associated with vibrations parallel to the bonds, $\omega_\parallel = k_B \Theta_\parallel/\hbar$
is the parallel Einstein angular frequency, $\mu$ is the reduced mass for an absorber-scatterer pair,
$k_{e,\parallel} = \mu \omega^2_\parallel$, $k_B$ is the Boltzmann's constant
and $C^*_{2,{\text{st}},\parallel}$ is the static or structural (independent of temperature)
contribution to the $\text{MSRD}_\parallel$. $C_3^*$ is  written as

\begin{equation}
C_3^* = \frac{3k_3 \hbar^6}{2\mu^3 k_B^4 \Theta_{\parallel}^4}
\biggl\{\Bigl[\coth{\bigl(\frac{\Theta_{\parallel}}{2T} \bigr)} \Bigr]^2 -1\biggr\} + C_{3,{\text{st}}}^* \,,
\label{eqc3}
\end{equation}

\noindent where $C_{3,{\text{st}}}^*$ is the static or structural (independent of temperature) contribution
to the asymmetry of $\varrho(r,T)$ and $k_3$ is related to the
cubic anharmonic term of $V(r)$ (see eq.~\ref{potencialefetivo}).
In a similar way, $\text{MSRD}_\perp=\langle (\Delta u_\perp)^2\rangle$ can be written as \cite{Vaccari,Schnohr}

\begin{equation}
C_\perp^*= \text{MSRD}_\perp = \langle (\Delta u_\perp)^2\rangle=
\frac{\hbar^2}{\mu k_B \Theta_{\perp}} \coth{\bigl(\frac{\Theta_{\perp}}{2T} \bigr)}  \,.
\label{eqmsrdperp}
\end{equation}

\noindent Here, $\Theta_{\perp}$ is the Einstein temperature associated with vibrations perpendicular to the
bonds. The temperatures $\Theta_{\perp}$ and $\Theta_{\parallel}$ should be the same only in isotropic materials.
If a crystalline sample
is under investigation, EXAFS measurements at some different temperatures furnish $C_1^*$, $C_2^*$ and
$C_3^*$ and, considering eqs. \ref{eqc2}--\ref{eqmsrdperp} together with eq.~\ref{eqrc} and $R_c$
obtained from XRD measurements, the quantities $\Theta_\parallel$, $\Theta_\perp$, $\text{MSRD}_\parallel$,
$\text{MSRD}_\perp$ and $k_3$ can in principle be obtained, furnishing information about anharmonicity, asymmetric
distribution functions, anisotropic vibrations and also, from $\Delta C_1^*$, thermal expansion, which is different
from the XRD thermal expansion due to the $\text{MSRD}_\perp$ in eq. \ref{cumulantec1}.
The problem now is how to obtain such information for an
amorphous sample. The first point is that the crystallographic distance $R_c$ should be substituted for
an amorphous XRD distance $R_a$, that is, from eq.~\ref{eqdefrc},

\begin{equation}
R_a = R_0 + \langle\Delta u_\parallel\rangle
\label{eqdefra}
\end{equation}

\noindent which should, in principle, be obtained by XRD. However, determination of average interatomic
distances for amorphous materials is not easy and, when it can be done, usually the error bars are large,
making the $\text{MSRD}_\perp$ obtained from inversion of eq. \ref{cumulantec1} unreliable. We
have developed a method to obtain all data above using only EXAFS measurements, which could also be used to
crystalline samples, but it needs several EXAFS measurements at different temperatures and on edges of both
atomic species (for a binary alloy) to work. We illustrate the method in sec. \ref{secresultados},
by investigating an amorphous Ni$_{46}$Ti$_{54}$ alloy.

\section{Experimental Details}
\label{secexperimental}

The Ni$_{46}$Ti$_{54}$ alloy was prepared by milling
Ni (Merck, purity $>$ 99.5 \%) and Ti (Alfa Aesar, purity $>$ 99.5 \%) crystalline powders under argon atmosphere
in a steel vial considering a ball to powder ratio of 5:1. The vial was mounted on a high energy Spex 8000 shaker mill
and the powders were milled for 12 h.

EXAFS measurements on Ni and Ti K edges were taken in the transmission mode at beam
line D04B-XAFS1 of the Brazilian Synchrotron Light Laboratory - LNLS. Three ionization chambers were used as detectors.
{\em a}-Ni$_{46}$Ti$_{54}$ samples were formed by placing the powder on a porous membrane (Millipore, 0.2 $\mu$m pore
size) and they were placed between the first and second chambers.
Crystalline Ni and Ti foils furnished by LNLS were used as energy references and were placed between the
second and third chambers. The beam size at the samples was 3 mm $\times$ 1 mm. The energy and average current of
the storage ring were 1.37 GeV and 190 mA, respectively. EXAFS data were acquired at 30, 100, 200 and 300 K on Ni K edge
and at 20 and 300 K on Ti K edge.
The raw EXAFS data were analyzed following standard procedures using ATHENA and ARTEMIS \cite{athena} programs.
Fourier transforms were performed considering Hanning window functions in the following ranges:
3.1--14.0 \AA$^{-1}$ (Ni K edge) and 3.5--12.9 \AA$^{-1}$ (Ti K edge) for the photoelectron momentum $k$ and
1.0--2.8~\AA\ (Ni K edge) and 2.0--3.3~\AA\ (Ti K edge) for the uncorrected phase radial distance $r$.
Amplitudes and phase shifts  were obtained from {\em ab initio} calculations using the spherical waves
method \cite{Rehr} and FEFF8.02 software. Each measurement was fitted simultaneously with multiple $k$
weightings of 1--3 to reduce correlations between the fitting parameters. 
It should be noted that using the above procedure on an EXAFS analysis,
the real cumulants $C_n^*$ are obtained \cite{Araujo,Schnohr}, not the effective ones.

\section{Results and Discussion}
\label{secresultados}

\begin{figure}[h]
\includegraphics{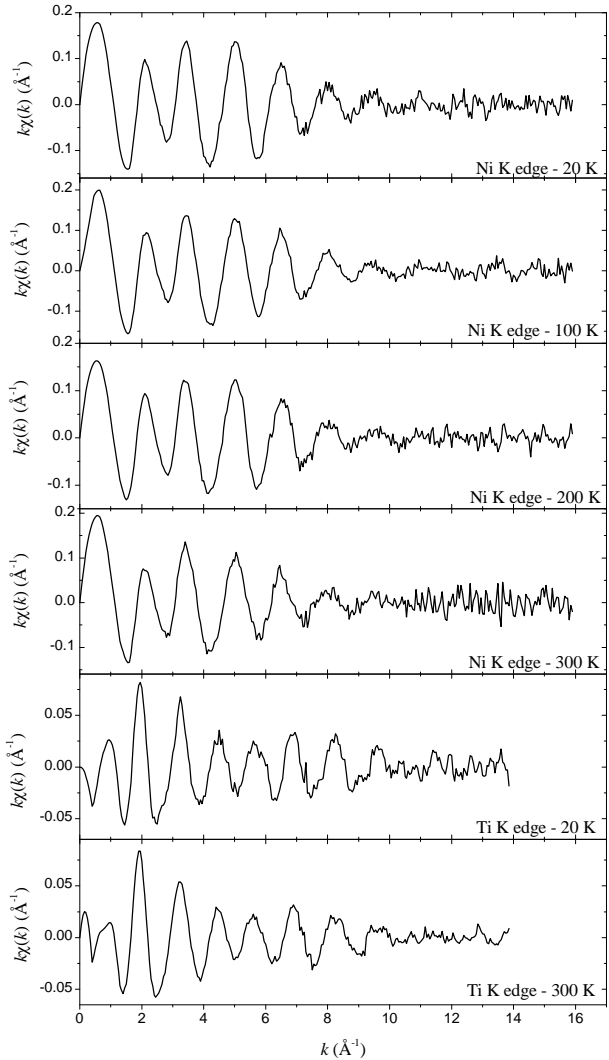}
\caption{\label{fig_chi_todas} EXAFS $k\chi(k)$ oscillations obtained on Ni and Ti K edges for {\em a}-Ni$_{46}$Ti$_{54}$.}
\end{figure}

\begin{figure}[h]
\includegraphics{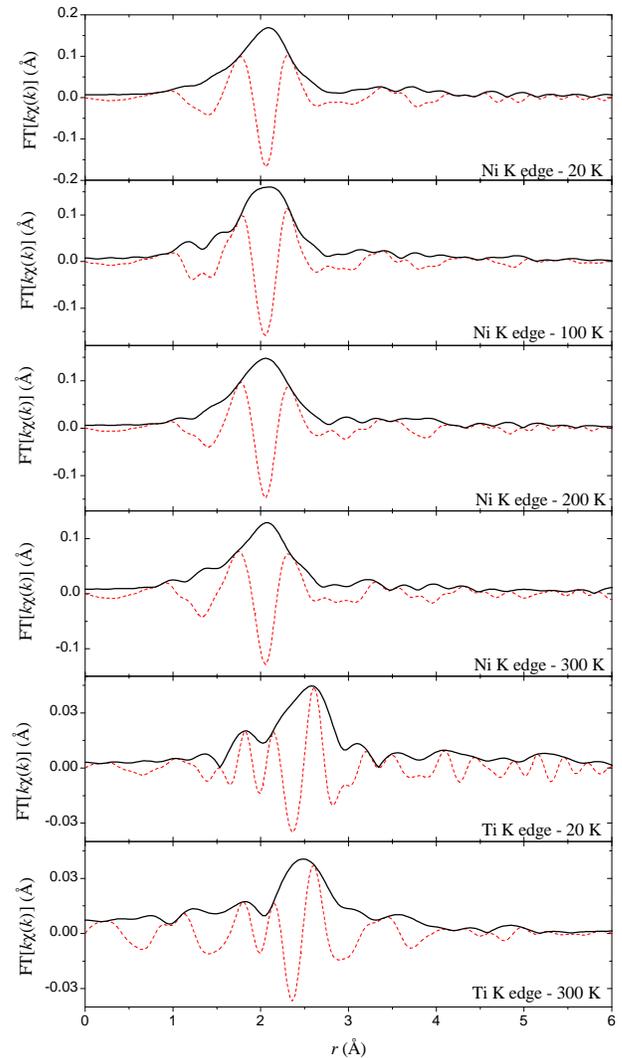}
\caption{\label{figtodasft} Magnitudes and imaginary parts of the Fourier transforms of the EXAFS $k\chi(k)$ data shown in 
fig.~\ref{fig_chi_todas} for {\em a}-Ni$_{46}$Ti$_{54}$.}
\end{figure}

Due to the fact that we had many EXAFS data at several temperatures an on Ni and Ti K edges, we could use several
constraints during the fits. Fig.~\ref{fig_chi_todas} shows the EXAFS $k\chi(k)$ oscillations obtained from the measurements, 
and fig.~\ref{figtodasft} shows the magnitudes and imaginary parts of their Fourier transforms. It is interesting to note 
that the  maxima of magnitudes and imaginary parts do not coincide, which is an indication of the asymmetry of 
the $g_{ij}(r)$ functions \cite{Koningsberger2}. Besides that, only the first shell is seen, as is expected for amorphous 
samples.

We made many tests considering different combinations of the experimental data and we
will discuss here in details two models, defined as

\begin{description}
\item[Model A] all experimental data were used without constraints related to MSRD$_\perp$.

\item[Model B] all experimental data were used together with the constraints related to MSRD$_\perp$.
\end{description}

\begin{figure}[h]
\begin{center}
\includegraphics[scale=0.5]{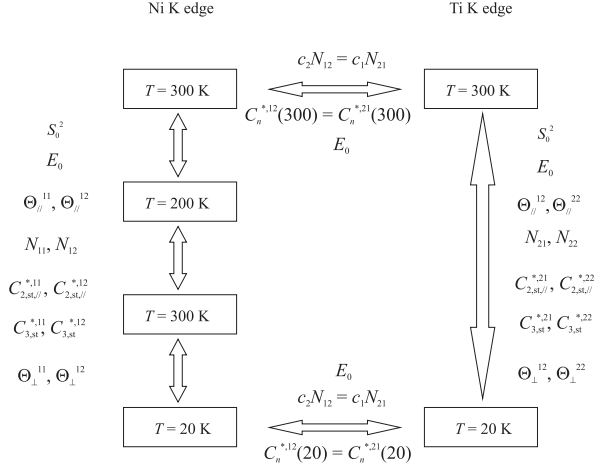}
\end{center}
\caption{\label{figdiagrama} Diagram showing the possible constraints that can be used in the EXAFS analyses.}
\end{figure}

To help the ``visualization" of the possible constraints that can be introduced during the fitting process,
fig.~\ref{figdiagrama} presents a diagram showing all relations used in the various EXAFS analyses we made. In the
diagram, $c_i$ is the concentration of atoms of type $i$, where $i=1$ for Ni and $i=2$ for Ti.

\begin{figure}[!]
\includegraphics{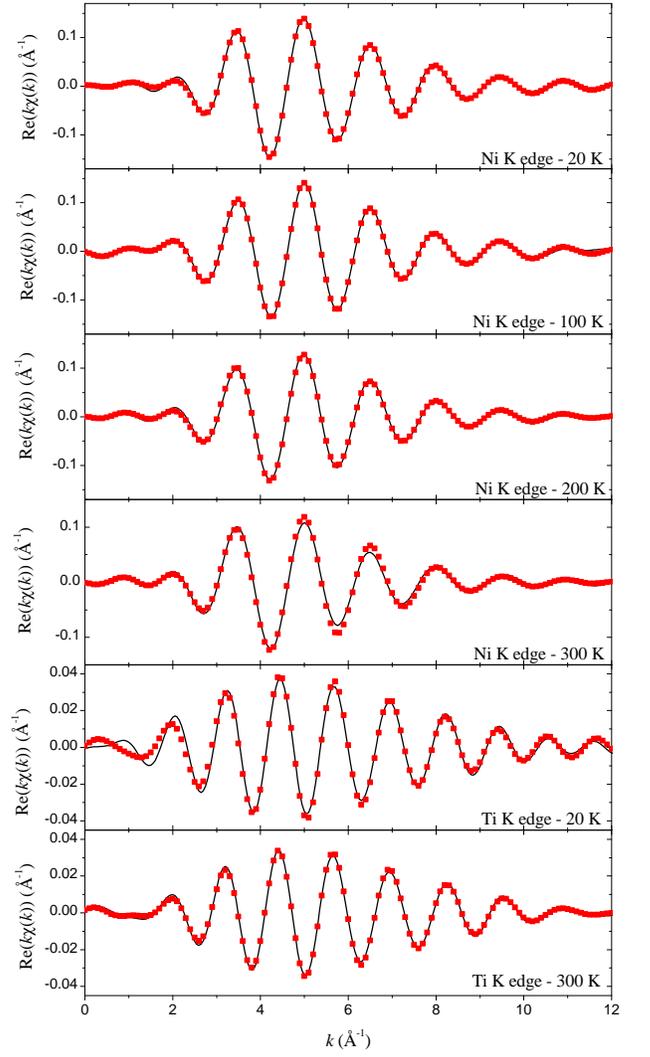}
\caption{\label{figchitodas} Real part of the Fourier filtered first shells of {\em a}-Ni$_{46}$Ti$_{54}$
(solid black lines) on Ni and Ti K edges at all temperatures investigated and their simulations (red squares).}
\end{figure}

\begin{figure}[!]
\includegraphics{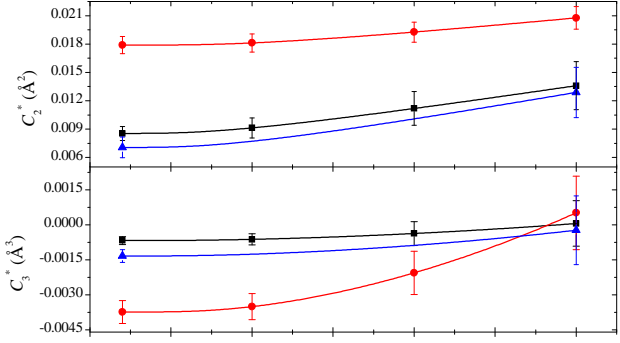}
\caption{\label{figc2c3} Cumulants $C_2^{*,ij}$ and $C_3^{*,ij}$ for {\em a}-Ni$_{46}$Ti$_{54}$ obtained for Ni-Ni
(black squares), Ni-Ti (red circles) and Ti-Ti (blue triangles) as functions of temperature considering
eqs.~\ref{eqc2} and~\ref{eqc3} respectively.}
\end{figure}

In Model A we did not consider any constraints for the first cumulant $C_1^*$ related to the MSRD$_\perp$, and
we did not consider perpendicular Einstein temperatures. All other constraints shown in fig.~\ref{figdiagrama}
were introduced in order to decrease correlations. The cumulants $C_2^{*,ij}$ and $C_3^{*,ij}$ associated with
each pair $ij$ were constrained to follow eqs. \ref{eqc2} and
\ref{eqc3}, respectively. Then, we obtained the threshold energy $E_0$,
$N_{ij}$ (average coordination numbers), $C^{*,ij}_{2,{\text{st}},\parallel}$ and
$C^{*,ij}_{3,{\text{st}}}$, $\Theta_\parallel^{ij}$, $k_3^{ij}$ and $C_1^{*,ij} = \langle r_{ij} \rangle$.
Figure~\ref{figchitodas} shows the real parts of the Fourier filtered first shells obtained on Ni and Ti K edges.
The agreement between the simulations and the experimental data is very good, and
it happens for all measurements on both edges at all temperatures. Tables~\ref{tab1}, \ref{tab2} and~\ref{tab3} present
the values obtained for the relevant quantities above considering model A, and fig.~\ref{figc2c3} shows the temperature
dependence of $C_2^{*,ij}$ and $C_3^{*,ij}$ (see eqs. \ref{eqc2} and \ref{eqc3}).

\begin{table}[!]
\begin{ruledtabular}
\caption{\label{tab1} Average coordination numbers and structural components of cumulants $C^*_2$ and $C^*_3$
obtained from the EXAFS fits shown in fig.~\ref{figchitodas} for {\em a}-Ni$_{46}$Ti$_{54}$ considering model A.}
\begin{tabular}{cccc}
Bond Type & $N$ & $C^*_{2,{\text{st}}}$ ($\times 10^{-3}$ \AA$^2$) &
$C^*_{3,{\text{st}}}$ ($\times 10^{-4}$ \AA$^3$)\\
Ni-Ni & $6.3  \pm 1.0$ & $5.3 \pm 0.2$ & $-0.72  \pm 0.01$ \\
Ni-Ti & $7.1 \pm 0.6$ & $15.1 \pm 0.8$ & $-4.2 \pm 0.4$ \\
Ti-Ti & $5.8 \pm 0.3$ & $3.1 \pm 0.4$ & $-1.4 \pm 0.2$  \\
\end{tabular}
\end{ruledtabular}
\end{table}

\begin{table}[!]
\begin{ruledtabular}
\caption{\label{tab2} Parallel Einstein temperatures, parallel effective harmonic spring constants, parallel Einstein
frequencies and cubic anharmonic force constants obtained from the EXAFS fits shown in fig.~\ref{figchitodas} for
{\em a}-Ni$_{46}$Ti$_{54}$ considering model A.}
\begin{tabular}{ccccc}
Bond Type & $\Theta_\parallel$ (K) & $k_{e,\parallel}$ (eV/\AA$^2$) & $\nu_\parallel$ (THz) & $k_3$ (eV/\AA$^3$)\\
Ni-Ni & $251 \pm 36$ & 3.3  & 5.2 & $6.8 \pm 1.5$\\
Ni-Ti & $326 \pm 12$ &  5.0 & 6.8 & $144 \pm 5$\\
Ti-Ti & $257 \pm 27$ & 2.8 & 5.4 & $6.5 \pm 2.3 $ \\
\end{tabular}
\end{ruledtabular}
\end{table}

\begin{table}[!]
\begin{ruledtabular}
\caption{\label{tab3} First cumulant $C_1^{*,ij}$ obtained from the EXAFS fits shown in fig.~\ref{figchitodas} for
{\em a}-Ni$_{46}$Ti$_{54}$ considering model A.}
\begin{tabular}{ccccc}
$T$ (K) & $C_1^{*,\text{Ni-Ni}}$ (\AA) & $C_1^{*,\text{Ni-Ti}}$ (\AA) & $C_1^{*,\text{Ti-Ti}}$ (\AA)\\
20 & $2.406 \pm 0.004$ & $ 2.447 \pm 0.006$  & $2.814 \pm 0.003$\\
100 & $2.413 \pm 0.003$ & $ 2.445 \pm 0.005$  & ---\\
200 & $2.418 \pm 0.005$ & $ 2.479 \pm 0.006$  & ---\\
300 & $2.455 \pm 0.011$ & $ 2.508 \pm 0.005$  & $2.852 \pm 0.002$\\
\end{tabular}
\end{ruledtabular}
\end{table}

From Table~\ref{tab1} and fig.~\ref{figc2c3} it can be seen that the contribution of the parallel structural disorder
$C_{2,\text{st},\parallel}^{*,ij}$ to the total MSRD$_\parallel$ is large, and it is the dominant factor even at 300 K for
Ni-Ti pairs. We believe this fact can be associated with the fact that the sample is amorphous and also with
fabrication technique used to produce the alloy. The third cumulant $C_3^{*,ij}$ indicates asymmetric distributions
$\varrho_{ij}(r,T)$ at low temperatures due mainly to the contribution of the structural part since
$C_{3,{\text{st}}}^{*,ij}$ is negative, but the asymmetry decreases with temperature. It is
interesting to note the thermal behaviour of $C_{3}^{*,\text{Ni-Ti} }$, which is
different from $C_{3}^{*,\text{Ni-Ni}}$ and $C_{3}^{*,\text{Ti-Ti}}$. This fact can be associated with
the parallel Einstein temperatures obtained, since homopolar pairs have similar values
and they are smaller than the value obtained for $\Theta_\parallel^{\text{Ni-Ti}}$ (see Table~\ref{tab2}),
and also with the anharmonicity of $V(r)$, indicated by the cubic anharmonic constants $k_3^{ij}$. The average interatomic
distances (first cumulant) $C_1^{*,ij}$ indicate a thermal expansion for Ni-Ni and Ti-Ti pairs and a negative (although
very small) thermal expansion for Ni-Ti pairs from 20 K to 100 K followed by a positive thermal expansion. From $\Delta
C_1^{*,ij}$ the thermal expansion coefficients $\alpha_{ij}(T)$ can be estimated, and that will be done latter.

\begin{figure}[h]
\includegraphics{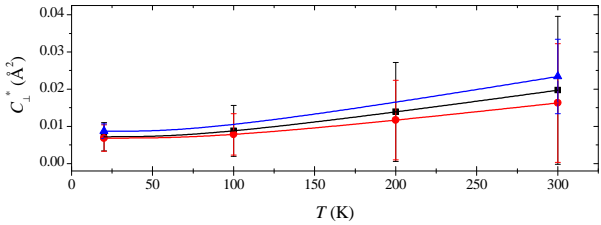}
\caption{\label{figcperp} Cumulants $C_\perp^{*,ij}$ for {\em a}-Ni$_{46}$Ti$_{54}$ obtained for Ni-Ni
(black squares), Ni-Ti (red circles) and Ti-Ti (blue triangles) as a function of temperature considering
eq.~\ref{eqmsrdperp}.}
\end{figure}

\begin{figure}[h]
\includegraphics{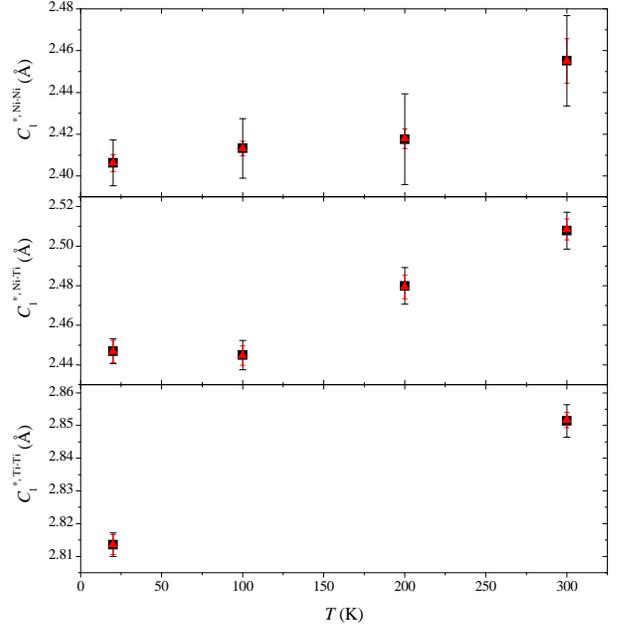}
\caption{\label{figc1} Comparison between cumulants $C_1^{*,ij}$ obtained from model A (red triangles) and
model B (black squares).}
\end{figure}

Now we can discuss model B. Since we had many constraints all structural values were obtained with enough accuracy to
proceed to the next step. We fixed all values shown in Tables~\ref{tab1}, \ref{tab2} and~\ref{tab3} except $C_1^{*,ij}$,
which were then constrained to follow eq.~\ref{cumulantec1} with
$\langle (\Delta u_\perp^{ij})^2\rangle$ given by eq. \ref{eqmsrdperp}.\
Since we knew $C_1^{*,ij}$ from model A, we could analyze the new
values obtained for $C_1^{*,ij}$ in order to avoid possible spurious results. The values obtained for the first cumulant
were almost the same, and we were able to find $\Theta_\perp^{ij}$, $\text{MSRD}_\perp^{ij}$,
$\langle\Delta u_\parallel^{ij} \rangle$, $R_0^{ij}$ and the corresponding amorphous distance $R_a^{ij}$ given by
eq.~\ref{eqdefra}. The fits obtained for model B are almost identical to those shown in Fig.~\ref{figchitodas} and
will not be shown.
Fig.~\ref{figcperp} shows the thermal behavior of $C^{*,ij}_{\perp}$ given by
eq.~\ref{eqmsrdperp} and Fig.~\ref{figc1} compares the values obtained for $C_1^{*,ij}$ using models A and B. As it can
be seen from Fig.~\ref{figc1} and also from Tables~\ref{tab3} and~\ref{tab4}, the values obtained from both models
are very similar. Table~\ref{tab5} shows other relevant structural parameters obtained.

\begin{table}[h]
\begin{ruledtabular}
\caption{\label{tab4} First cumulant $C_1^{*,ij}$ obtained from the EXAFS fits shown in Fig.~\ref{figchitodas} for
{\em a}-Ni$_{46}$Ti$_{54}$ considering model B.}
\begin{tabular}{ccccc}
$T$ (K) & $C_1^{*,\text{Ni-Ni}}$ (\AA) & $C_1^{*,\text{Ni-Ti}}$ (\AA) & $C_1^{*,\text{Ti-Ti}}$ (\AA)\\
20 & $2.406 \pm 0.011$ & $ 2.447 \pm 0.006$  & $2.814 \pm 0.004$\\
100 & $2.413 \pm 0.014$ & $ 2.445 \pm 0.007$  & ---\\
200 & $2.418 \pm 0.021$ & $ 2.480 \pm 0.009$  & ---\\
300 & $2.455 \pm 0.022$ & $ 2.508 \pm 0.009$  & $2.851 \pm 0.005$\\
\end{tabular}
\end{ruledtabular}
\end{table}

\begin{table}[!]
\begin{ruledtabular}
\caption{\label{tab5} Rest distances $R_0^{ij}$, perpendicular Einstein temperatures, perpendicular effective harmonic
spring constants and perpendicular Einstein frequencies  obtained from the EXAFS fits for {\em a}-Ni$_{46}$Ti$_{54}$
considering model B.}
\begin{tabular}{ccccc}
Bond Type & $R_0^{ij}$ (\AA) & $\Theta_\perp$ (K) & $k_{e,\perp}$ (eV/\AA$^2$) & $\nu_\perp$ (THz) \\
Ni-Ni & $2.400 \pm 0.006$ & $230 \pm 121$ & 2.7  & 4.8 \\
Ni-Ti & $2.438 \pm 0.002$ & $269 \pm 140$ &  3.4 & 5.6 \\
Ti-Ti & $2.810 \pm 0.001$ & $233 \pm 52$ & 2.3 & 4.9 \\
\end{tabular}
\end{ruledtabular}
\end{table}

\begin{figure}[h]
\includegraphics{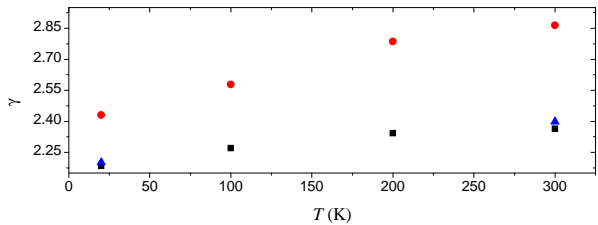}
\caption{\label{figgama} Ratio $\gamma_{ij} = C_\perp^{*,ij}/C_{2,\parallel,T}^{*,ij}$ obtained from EXAFS
for Ni-Ni (black squares), Ni-Ti (red circles) and Ti-Ti (blue triangles) pairs in {\em a}-Ni$_{46}$Ti$_{54}$.}
\end{figure}

Due to similar $\Theta_\perp^{ij}$, the thermal behavior of $C^{*,ij}_\perp$ shown in
Fig.~\ref{figcperp} is similar. The perpendicular Einstein temperatures are smaller than the corresponding
$\Theta_{\parallel}^{ij}$, the difference being larger
for Ni-Ti pairs, indicating the presence of
vibrational anisotropy for all pairs specially for Ni-Ti ones. The corresponding force constants and frequencies
indicate a loosening of the perpendicular bond strengths when compared
to the parallel ones. So, bending vibrational modes are more easily excited than
stretching modes.
Fig.~\ref{figgama} shows the ratio $\gamma_{ij} = C_\perp^{*,ij}/C_{2,\parallel,T}^{*,ij}$
obtained using eq.~\ref{eqc2} and eq.~\ref{eqmsrdperp}. The values
found are greater than 2, indicating the vibrational anisotropy, which is higher for the Ni-Ti pairs.

The values obtained for the first cumulants $C_1^{*,ij}$ shown in Fig.~\ref{figc1} and Tables~\ref{tab3} and~\ref{tab4}
indicate a thermal expansion for Ni-Ni and Ti-Ti pairs for all temperatures and for Ni-Ti pairs after 100 K.
However, these pairs seem to have a negative thermal expansion from 20 K to 100 K. Considering
eqs.~\ref{cumulantec1}, \ref{rparaperp} and \ref{eqmsrdperp}, the perpendicular contribution $r_\perp^{ij}$
to $C_1^{*,ij}$ can be calculated and it is shown, together with the
parallel contribution $r_\parallel^{ij}$, in Table~\ref{tab6}.

\begin{table*}[!]
\begin{ruledtabular}
\caption{\label{tab6} First cumulant $C_1^{*,ij}$ obtained from the EXAFS fits shown in Fig.~\ref{figchitodas} for
{\em a}-Ni$_{46}$Ti$_{54}$ considering model B.}
\begin{tabular}{ccccccc}
$T$ (K) & $r_\parallel^{\text{Ni-Ni}}$ (\AA) & $r_\perp^{\text{Ni-Ni}}$ (\AA) &
$r_\parallel^{\text{Ni-Ti}}$  (\AA) & $r_\perp^{\text{Ni-Ti}}$ (\AA)
& $r_\parallel^{\text{Ti-Ti}}$  (\AA) & $r_\perp^{\text{Ti-Ti}}$ (\AA)\\
20 & $0.004 \pm 0.005$ & $ 0.0015 \pm 0.0008$  & $0.007 \pm 0.004$ & $0.0014 \pm 0.0007$
& $0.002 \pm 0.002$ & $0.0015 \pm 0.0003$\\
100 & $0.011 \pm 0.007$ & $ 0.002 \pm 0.001$  & $0.005 \pm 0.004$& $0.002 \pm 0.001$
& --- & $0.0019 \pm 0.0006$\\
200 & $0.014 \pm 0.013$ & $ 0.003 \pm 0.003$  & $0.039 \pm 0.005$ &  $0.002 \pm 0.002$
& --- & $0.003 \pm 0.001$\\
300 & $0.051 \pm 0.012$ & $ 0.004 \pm 0.004$  & $0.066 \pm 0.004$ & $0.003 \pm 0.003$
& $0.037 \pm 0.002$ & $0.004 \pm 0.002$\\
\end{tabular}
\end{ruledtabular}
\end{table*}

From Table~\ref{tab6} it can be seen that the perpendicular contribution $r_\perp^{ij}$ associated with
the rigid shift of the potential minimum is smaller than the term $r_\parallel^{ij}$, which is related to the
anharmonicity and to the shape of the effective potential, and their difference increases as the temperature is
raised. This fact suggests that the thermal expansion in {\em a}-Ni$_{46}$Ti$_{54}$ is caused mainly by changes
in the shape of the potential and the rigid shift of the potential minimum is a secondary effect.

The thermal expansion coefficient $\alpha$ can be written as

\begin{equation}
\alpha(T) = \frac{1}{r} \frac{dr}{dt} = \frac{1}{C_1^*} \frac{dC_1^*}{dt} =
\frac{1}{C_1^*} \frac{dr_\parallel}{dt} + \frac{1}{C_1^*} \frac{dr_\perp}{dt} =
\alpha_\parallel + \alpha_\perp
\end{equation}

The contribution $\alpha_\perp$ to the thermal expansion associated with perpendicular vibrations can be found
exactly  and typical values are $\alpha_\perp^{\text{Ni-Ni}} = 5.1 \times 10^{-6}$ K$^{-1}$ at 300 K and
$\alpha_\perp^{\text{Ni-Ti}} = 2.4 \times 10^{-6}$ K$^{-1}$ at 100 K (the largest and the smallest values, respectively).
The total thermal expansion can be estimated from $\Delta C_1^{*,ij}$
and, for Ni-Ti pairs at 100 K, it is negative and has the value
$\alpha^{\text{Ni-Ti}} = -9.6 \times 10^{-6}$ K$^{-1}$.
All other values are positive and increase with temperature, ranging from
$\alpha^{\text{Ni-Ni}} = 3.6 \times 10^{-5}$ K$^{-1}$, at 100 K, to
$\alpha^{\text{Ni-Ni}} = 1.5 \times 10^{-4}$ K$^{-1}$, at 300 K. The contribution $\alpha_\perp$ to the total
thermal expansion is always much smaller than $\alpha_\parallel$ except for Ni-Ti pairs at 100 K, indicating that
the anharmonicity of the effective potencial is the dominant effect related to thermal expansion for
the amorphous Ni$_{46}$Ti$_{54}$ alloy studied.

\section{Conclusion}
\label{secconclusoes}

We investigated the structure, anharmonicity, asymmetry of pair distribution functions, vibrational anisotropy
and thermal expansion of an amorphous Ni$_{46}$Ti$_{54}$ alloy produced by mechanical alloying considering EXAFS
data only. The detailed study described here was only possible due to the
experimental data available on Ni and Ti K edges and at several temperatures, which allowed us to use many
constraints to obtain reliable values for the cumulants $C_1^*$, $C_2^*$ and $C_3^*$,
making it possible to extract the perpendicular contributions to the MSRD and $C_1^{*,ij}$. The method should also
work for crystalline samples, but we believe that EXAFS measurements on edges of all atomic species present in the sample
should be used. This procedure may be the only way to extract such structural information for amorphous samples, so
more tests on other alloys would be important.

Concerning the Ni$_{46}$Ti$_{54}$ alloy, it presents asymmetric $\varrho_{ij}(r,T)$ functions at low temperatures due to
the structural contribution but the asymmetry decreases with temperature until 300 K. After that, the asymmetry incrases
again. There is vibrational anisotropy mainly for Ni-Ti pairs, indicating that bending modes are looser than stretching
ones, and Ni-Ti pairs exhibits a negative thermal expansion around 100 K. It would be very interesting to study
the crystalline shape memory NiTi phase in order to associate these properties with the mechanical properties of this
alloy.

\acknowledgments

We would like to thank the Brazilian agency CNPq for financial support. This study was also partially
supported by LNLS (proposal XAFS1 4367/05).


\end{document}